\documentclass[showpacs,preprint,prd,nofootinbib]{revtex4}

\usepackage{amssymb}
\usepackage{amsmath}
\usepackage{graphicx}  

\usepackage{color}

\usepackage{soul}

\begin{document}

\title{Signatures of electromagnetic and sound radiation produced by gravitational waves in core-collapse supernovae}

\author{Preston Jones}
\email{Preston.Jones1@erau.edu}
\author{Pragati Pradhan}
\email{pradhanp@erau.edu}
\affiliation{Embry Riddle Aeronautical University, Prescott, AZ 86301}
\author{Douglas Singleton}
\email{dougs@mail.fresnostate.edu}
\affiliation{California State University Fresno, Fresno, CA 93740}

\date{\today}

\begin{abstract}
This paper presents the scenario that gravitational waves, generated in core-collapse of a pre-supernova star, can produce both electromagnetic radiation and sound radiation as gravitational waves propagate outward from the collapsing core. While the energy of this co-produced electromagnetic and sound radiation is orders of magnitude smaller than the initiating gravitational radiation, the power may be sufficient to re-ignite fusion outside the collapsing core.  The non-equilibrium re-ignition of fusion, in roughly the same time frame as the strongest neutrino emissions, would change the configuration of the pre-supernova star and subsequently the ejecta and the evolution of the stellar expansion of the supernova remnant. Although the co-produced electromagnetic or sound radiation could not contribute directly to the supernova explosion, the associated non-equilibrium re-ignition of fusion would alter the state outside the core leaving an observable signature in the ejecta of the supernova remnant. The aim of this paper is to argue that including this hypothesized co-produced radiation in computational models of core collapse supernovae would contribute to the evolution of the stellar expansion and consequently should be observable in the supernovae remnant, providing a confirmation of the conversion processes for gravitational radiation to electromagnetic and sound radiation.
\end{abstract}


\maketitle

\section{Introduction}

The observation of gravitational waves has opened a new window on the Universe and our understanding of astrophysics. The coordinated international, multi-messenger observation of a kilonova \cite{ligo} is a testament to the incredible potential for the future of astrophysics. Equally tantalizing is the unresolved signal above 300 Hz in the kilonova gravitational wave observation and the limitations this presents in understanding astronomical events. The higher frequency detection of gravitational waves above 300 Hz is especially important for multi-messenger studies, via gravitational waves, electromagnetic waves and neutrinos, of core-collapse supernovae (CCSNe). However, the explosive mechanisms for core-collapse supernova and the evolution of the supernova remnant are still not completely understood \cite{Janka17}. This leaves open the possibility that the physics in the current models might not be complete - there may be interesting additional physical mechanisms involved. Recent models of the evolution of the CCSNe remnants describe potential signatures of such a mechanism, e.g., ejecta associated with $\beta$  decay  \cite{Gabler21}. The aim of this paper is to describe comparable and previously unconsidered physical mechanisms and propose the inclusion of these mechanisms in future models of CCSNe and CCSNe Remnants (SNR). Specifically, the production of electromagnetic and sound (acoustic) radiation by gravitational waves generated by the collapsing core.

The possibility that neutrino-driven explosion mechanisms are not the only ``input physics" important in core-collapse supernova was previously considered in \cite{Janka17} for models of CCSNe and in \cite{Gabler21} for models of SNR. Our focus here is on the ``input physics'' for the SNR computational models \cite{Gabler21}. Specifically, potential signatures of production of electromagnetic and sound radiation produced by gravitational waves (gw). These models continue as the star expands and asymmetries develop in the ejecta following the collapse and explosion. As the computational models continue to evolve for days, weeks, and months the ``input physics" will contribute to signatures in the asymmetries of the ejecta of the expanding star (SNR). These signatures should be observable e.g., in the x-ray spectra of the remnant ejecta of the SNR as discussed in Sections \ref{remnant} and \ref{observation}. One possible piece of previously unconsidered  ``input physics" is the mechanism of conversion of gravitational waves to electromagnetic radiation or sound radiation coincident with the core-collapse. At the particle or Feynman diagram level the process of graviton-photon scattering, photon creation from gravitons, was studied at tree level for photons by Skobelev \cite{Skobelev75,Sawyer20}. Around the same time Gibbons \cite{gibbons} studied the possibility for gravitational wave backgrounds (i.e. many gravitons) to create other classical wave fields (i.e. many photons/many phonons). Gibbons found the following restriction on this conversion process: ``Indeed since a `graviton' presumably in some sense carries light-like momentum the creation of one or more particles with time-like or light-like momentum would violate the conservation of momentum {\em unless the created particles were massless} and precisely aligned with the momentum of the graviton", with the emphasis added. For a long time Gibbons' result has been taken to imply that production of an electromagnetic wave (photons) or sounds wave (phonons) from a gravitational wave background (gravitons) was forbidden. However, combining the results of Gibbons and Skobelev show that the production of electromagnetic waves from gravitational waves is possible, so long as the production is massless and the wave travels in the same direction as the generating gravitational wave. The Feynman diagram calculation of Skobelev indicated that this conversion of gravitons to photons was so small as to make it unlikely that these processes would be of consequence in any astrophysical process  \cite{Skobelev75}, at least when looking at single photons and single gravitons.

Based on our recent work on the semi-classical conversion of gravitational waves to electromagnetic radiation \cite{Jones15,Jones16,Jones17,Jones18,Gretarsson18}, we demonstrate that the electromagnetic energy produced is a small fraction of the initiating gravitational radiation energy, and is far too small to contribute directly to the explosion. On the other hand we demonstrate that the power of the co-produced elctromagnetic radiation is sufficient to re-ignite fusion in the outer core layers. The production of gravitational radiation at core collapse will occur in roughly the same time frame as the emission of neutrinos. Fusion initiated by the counterpart electromagnetic radiation would alter the state of the outer core of the pre-explosion star and contribute to the evolution of the expanding star. Since models of CCSNe and SNR are highly dependent on the initial state of the star in the model, the counterpart electromagnetic radiation would be expected to meaningfully impact the modeling of the supernovae and the later stages as a supernova remnant.

After studying the production of electromagnetic radiation from gravitational waves we investigate a previously unconsidered mechanism whereby gravitational radiation is converted into sonic radiation, sounds waves or phonons. This new mechanism of gravitons converting to phonons is based on work \cite{Visser10,Sabin14} which examined gravitational waves interacting with fluids. This graviton to phonon conversion is enhanced by a factor of the ratio of the speed of light to the speed of sound in the fluid compared to the graviton to photon process. Observations of SNR consistent with asymmetries and other structures in the evolving ejecta, associated with the production of electromagnetic and sound radiation, would provide evidence of these previously unconsidered physical mechanism.  \\

\section{Production of electromagnetic radiation from gravitational waves} \label{electromagnetic}

We first give a rough, first order calculation of how a gravitational wave (i.e. many gravitons) can give rise to an electromagnetic wave (i.e. many photons). The aim is to show that if there is enough energy in the gravitational wave (i.e. if there are a very large number of gravitons) this can compensate for the extreme smallness of the single Feynman diagram process of gravitons going to photons, $g+g \to \gamma + \gamma$, and give rise to a physically significant amount of vacuum produced electromagnetic radiation and energy. While the outward propagating gravitational wave interacts only very weakly with the matter of the progenitor star, the co-produced electromagnetic radiation would be completely absorbed.

The luminosity of the electromagnetic radiation, vacuum produced by the gravitational wave background, has been shown to be orders of magnitude less than the gravitational wave luminosity  \cite{Jones15,Jones16,Jones17,Jones18,Gretarsson18}. However, if the gravitational wave luminosity is very large, then the power radiated in the electromagnetic counterpart radiation can be large enough to be astrophysically significant, especially since electromagnetic radiation interacts much more strongly with ordinary matter as compared with gravitational radiation. In particular we will look at the possible impact this conversion of gravitational wave energy to electromagnetic wave energy might have on the core-collapse supernovae (CCSNe) and the evolution of the SNR. In a CCSNe it is widely accepted that neutrinos play the main role in driving the supernova explosion. However, numerical simulations which model supernova, and which incorporate only neutrinos as the mechanism behind the explosion and evolution are often models of the CCSNe, especially for larger mass progenitor stars \cite{Janka17}. By carrying out a rough calculation we show that it is possible that electromagnetic radiation produced by gravitational waves from the core collapse could significantly alter the state of the pre-explosion star and the subsequent processes of evolution of the SNR.

We present a simple, order of magnitude calculation of the electromagnetic radiation produced by a gravitational wave background. The equation of motion for the electromagnetic field will be solved for the curved space-time of a gravitational plane wave. We start with the Lagrange density for an electromagnetic field in the Lorenz gauge which is $\mathcal{L}_{em}  =  - \frac{1}{2}\partial _\mu  A_\nu  \partial ^\mu  A^\nu$. The vector potential can be written as $A_\mu  \left( {k,\lambda ,x} \right) = \epsilon_{\mu} ^{(\lambda)} \phi ^{(\lambda)} \left( {k ,x} \right)$ where $\epsilon_{\mu} ^{(\lambda)}$ is the polarization. Using this vector potential in $\mathcal{L}_{em}$ leads to a scalar-like Lagrange density $\mathcal{L}_{em}  =  - \partial _\mu  \varphi ^* \partial ^\mu  \varphi$, where $\varphi = \frac{1}{\sqrt{2}} \left( \phi^{(1)} + i \phi^{(2)} \right) $. Thus the electromagnetic Lagrange density has been reduced to the form of a scalar field Lagrange density.  Placing this scalar field, $\varphi (x)$, in a curved space-time background leads to the field equation

\begin{equation}
\frac{1}{{\sqrt{-g}}} \partial_{\mu} \sqrt{-g} g^{\mu \nu} \partial_{\nu}\varphi = 0.
\label{eomvarphiA}
\end{equation}

\noindent This is the massless Klein-Gordon equation in a curved background with metric $g^{\mu \nu}$. We will crudely approximate the metric for the gravitational wave propagating out from the core to have the plane wave form

\begin{equation}
ds^2 = g_{\mu \nu} dx^{\mu} dx^{\nu} = -dt^2 + dz^2 + a^2 dx^2 + b^2 dy^2. 
\label{GWmetric}
\end{equation}

\noindent We have set $c=1$ and taken the metric to be oscillatory with $a (u) =1 + \varepsilon (u) $ and $b (u) = 1 - \varepsilon(u)$ where $\varepsilon(u) =h e^{iku}$. Here the gravitational wave strain amplitude is $h$, the gravitational wave number is $k$, and $u=z-t$ is the standard light front coordinate for waves moving in the $+z$ direction. Substituting the metric \eqref{GWmetric} into \eqref{eomvarphiA} gives the field equation for the scalar field (see \cite{Jones16,Jones18} for details)

\begin{equation}
 \left( {4 F(u) \partial _u \partial _v  - 4ikG(u)\,\partial _v  + H(u) (\partial _x^2  + \partial _y^2) }\right)\varphi  = 0,
\label{eomvarphi}
\end{equation}

\noindent where $F\left( {u} \right)  \equiv  \left( {1 - 2h ^2 e^{2iku}  + h ^4 e^{4iku} } \right)$,  
$G\left( {u} \right)  \equiv  \left( {h ^2 e^{2iku}  - h ^4 e^{4iku} } \right)$, and $H(u) \equiv \left( 1 + h_+^2 e^{2iku} \right)$. The solution to \eqref{eomvarphi} can be shown \cite{Jones16,Jones18} to be 

\begin{equation}
\varphi  = A\frac{{e^{\frac{\lambda }{k}} \left( {1 - h^2 e^{2iku} } \right)^{\frac{1}{2}\left( {\frac{\lambda }{k} - 1} \right)} }}{{e^{\frac{\lambda }{{k\left( {1 - h^2 e^{2iku} } \right)}}} }}e^{ - i\lambda u} e^{ip_v v} e^{ipx} e^{ipy}  + B.
\label{Sfield}
\end{equation} \

\noindent $A$ and $B$ are normalization constants, $p, p_v$ are $\varphi$-field momenta in the $x,y$ and $v=z+t$ directions and $\lambda =\frac{p^2}{2 p_v}$. The normalization constants in (\ref{Sfield}) are set to $A = - B = 1$, consistent with the normalization of the Newman-Penrose scalars \cite{Teukolsky73} and the required vacuum solution for the fields in the absence of a gravitational wave, $h = 0$. We now take the vacuum limit of the scalar field in the presence of the gravitational wave i.e. $h \ne 0$. We do this by taking the $\varphi$-field momenta to zero ($p, p_v, \lambda \to 0$). In this vacuum limit of $\varphi$, we find that the scalar field does not go to zero but rather has the outgoing wave solution \cite{Jones16,Jones18}

\begin{eqnarray}
\varphi \left( {t,z} \right) &=& \left( {1 - h^2 e^{2ik\left( {z - t} \right)} } \right)^{ - \frac{1}{2}}  - 1 \nonumber \\
&\approx& \frac{1}{2}h^2 e^{2ik(z - t)}  + \frac{3}{8}h^4 e^{4ik(z - t)} . 
\label{OutState}
\end{eqnarray}
In the last line we assumed $h$ to be small. 

The relationship between the gravitational and electromagnetic luminosity or flux was developed in previous work \cite{Jones17} using the Newman-Penrose formalism  \cite{Teukolsky73}.  The luminosity or flux ratio for the gravitational and electromagnetic radiation in terms of the NP scalars is $\frac{d E_{em}/dt}{d E_{gw}/dt} = \frac{{\left( {\frac{1}{{4\pi }}\left| {\Phi _2 } \right|^2 } \right)}}{{\left( {\frac{1}{{16\pi k^2 }}\left| {\Psi _4 } \right|^2 } \right)}} $, where $\Phi_2$ is the electromagnetic NP scalar and $\Psi_4$ is the gravitational Newman-Penrose scalar. In \cite{Jones17} it was shown that the scalar field from (\ref{OutState}) and the metric given in (\ref{GWmetric}) led to $\left| \Phi_2 \right|^2 = 2 k^2 h^4$ and $\left| \Psi_4 \right|^2 = 4 k^4 h^2$ which resulted in $\frac{{\dot E_{em} }}{{\dot E_{gw} }} = 2h^2 \to F_{em} = 2 h^2 F_{gw}$, with $F_{em}$ and $F_{gw}$ being the electromagnetic and gravitational fluxes respectively. The gravitational wave flux associated with the metric \eqref{GWmetric} is $F_{gw}  = \frac{{c^3 }}{{16\pi G}}\left| {\dot \varepsilon } \right|^2 $ and $\left| {\dot \varepsilon } \right|^2 = h^2 \omega^2$. Collecting all these results gives the electromagnetic flux produced from the gravitational wave, in terms of the known plane wave gravitational parameters as

\begin{equation}
F_{em} =  \frac{ \pi {c^3}}{{2 G}} h^4 f^2 ,
\label{emFlux}
\end{equation}

\noindent where $f = \frac{\omega}{2 \pi}$ is the gravitational wave frequency.

\section{Counterpart production power in CCSN\lowercase{e}}

The power for the counterpart production of electromagnetic radiation from \eqref{emFlux} is

\begin{equation}
\dot E_{em} \left( r \right) = F_{em} 4\pi r^2  = 4\pi C \frac{{\tilde A^4 }}{{r^2 }}f^2 ,
\label{emPowerA}
\end{equation}

\noindent where $\tilde A = \left\langle {A_{20}^{E2} } \right\rangle $ is the mean quadrupole wave amplitude \cite{Muller04} which is related to the strain amplitude via $h = \frac{\tilde A}{r}$ and the multiplying constant $C =  \frac{ \pi {c^3}}{{2 G}} = {6 \times 10^{35} ~ \frac{{{\rm{Ws}}^{\rm{2}} }}{{{\rm{m}}^{\rm{2}} }}} = {6 \times 10^{42} ~ \frac{{{\rm{erg  \cdot s}} }}{{{\rm{m}}^{\rm{2}} }}} $. The total production of electromagnetic power from the gravitational wave propagating through a spatial region of one wavelength, $\lambda = \Delta r$, can be approximated from \eqref{emPowerA} by

\begin{equation}
\Delta \dot E_{em} \left( r \right) = \frac{{4\pi C }}{\lambda }\frac{{\tilde A^4 }}{{r^2 }}f^2 \Delta r ,
\label{SlicePower}
\end{equation}

\noindent which is then the power generated in a region equal to one wavelength. The power developed in any region greater than $\Delta r $, outside the inner core, will depend on the number of gravitational wave cycles in that region. The average number of cycles of the gravitational wave outside the inner core is, $\delta =  \Delta t f = \frac{{c\Delta t}}{\lambda }$, where $\Delta t$ is the duration of the gravitational wave. The total electromagnetic power produced is then found by going to the limit and integrating over the production distance,

\begin{equation}
\dot E^{tot} _{em}  = \delta \frac{{4\pi C }}{\lambda }\tilde A^4 f^2 \int\limits_{r_1}^{r_2} {\frac{{dr}}{{r^2 }}} .
\label{TotalPowerFinite}
\end{equation}

\noindent Assuming the production starts at a radius near the core $r_0 = r_1 $ and ends well outside the core $r_2 \gg r_0 $, or letting $r_2 \to \infty $, the total electromagnetic power is approximately 

\begin{equation}
\dot E^{tot} _{em}  \simeq \delta \frac{{4\pi C }}{c }\frac{{\tilde A^4 }}{{r_0 }}f^3 .
\label{TotalPowerIntB}
\end{equation}

The electromagnetic power in \eqref{TotalPowerIntB} is a rough estimate for the production from a spherically outward propagating gravitational wave. Here our interest is specifically in core-collapse proceeding a star's explosion and we make some estimates of $\dot E^{tot} _{em}$ for this case. Nominally the values for the the gravitational wave produced by core-collapse can be taken \cite{Muller04} to have duration of $\Delta t \sim 100 ~ {\rm{ms}}$, a mean quadrupole amplitude $\tilde A \sim 50 ~ {\rm{cm}}$, a frequency $f \sim 1 ~ {\rm{kHz}}$, $\delta \sim 100$, and an inner core radius of $r_0  \sim 10 ~{\rm{km}}$. Substituting these values into \eqref{TotalPowerIntB}, the total power is $\dot E^{tot} _{em}  \sim 10^{41} ~ \frac{{\rm{erg}}}{{\rm{s}}}$ and the total energy produced is estimated as $E^{tot} _{em} = \dot E^{tot} _{em} \Delta t = 10^{40} ~ \rm{erg}$ using $\Delta t \sim 100 ~ {\rm{ms}}$. This $10^{40} ~ \rm{erg}$ energy of the electromagnetic radiation, co-produced by the gravitational wave, is orders of magnitude too small to contribute directly to CCSNe shock recovery and explosion; this would require energies on the order of $10^{51} ~ {\rm{ erg}}$  \cite{Janka07,Janka12A}. While the energy of the counterpart electromagnetic radiation produced by the gravitational wave is far too small to contribute directly to shock recovery, the power of the electromagnetic production is still quite large, and is of an order of magnitude for re-ignition of fusion in the star's layers outside the iron core. This non-equilibrium ignition of fusion could contribute meaningfully to the evolution of the expanding star and the signature of this re-ignition should be observable in the SNR.

The fusion power equilibrium rate \cite{Hix99} is on the order of $10^{38} ~ \frac{{{\rm{erg}}}}{{\rm{s}}}$ and the above estimates show that the power of the co-produced electromagnetic radiation can exceed this value. If the power of the co-produced electromagnetic radiation is to re-ignite fusion, it must be outside the collapsing core at a radius on the order of $100 ~ {\rm{km}}$ \cite{Janka07,Janka12} i.e. outside the iron core and in the silicon layer. Substituting $r_0 \approx 100$ km into \eqref{TotalPowerIntB} the electromagnetic power outside the collapsing core is $10^{40} ~ \frac{{{\rm{erg}}}}{{\rm{s}}}$. Thus this co-produced electromagnetic power, coming from the gravitational wave, could re-ignite fusion outside the inner core. Under the proper conditions this could result in non-equilibrium fusion processes which could contribute locally to imbalance in the dynamical pressure of the neutrino mechanism \cite{Ott16} and potentially observable asymmetries in the SNR. Taking into account that the power produced from \eqref{TotalPowerIntB} is inversely proportional to the production radius, we find that the fusion equilibrium power would be exceeded out to a radius of $1,000 ~ {\rm{km}}$ which is well outside the collapsing core. These estimates all point to the ability of the co-produced electromagnetic radiation to re-ignite fusion burning. The timing of this hypothesized re-ignition of fusion is crucial if it is to play a role in CCSNe and in SNR. The re-ignition would occur in roughly the same time frame as the neutrino emission \cite{Muller04}. Such a re-ignition of silicon burning could contribute energy to the the expansion of the star as great as 10\% of the total supernova energy \cite{Janka17}.

One potential criticism of the conversion of gravitational waves to electromagnetic radiation presented in section II, is that the analysis depended crucially on the photons being massless. In fact since the gravitational waves are moving through a plasma, and since photons develop an effective mass inside a plasma given by (with factors of $\hbar$ and $c$ restored)

\begin{equation}
\label{photon-mass}
m_\gamma = \frac{4 \pi \hbar}{c^2} \sqrt{\frac{n q^2}{m}} = \hbar \omega_{plasma}~,
\end{equation}

\noindent where $n$, $q$ and $m$ are the number density, charge and mass of the component of the plasma. If there are several different components $\frac{n q^2}{m}$ would be replaced by $\sum _{i=1} ^n \frac{n_i q_i^2}{m_i}$, where one simply sums over the $n$ different components. If photons in plasma have a real, effective mass then the gravitational wave production of electromagnetic radiation/photons, described in section II would not occur. However, to understand the nature of the effective photon mass in \eqref{photon-mass} in more detail we write down the electromagnetic wave number in a plasma as

\begin{equation}
    \label{wave-number}
    k= \frac{2 \pi}{\lambda} \sqrt{1 - \frac{\omega_{plasma}^2}{\omega ^2}}~.
\end{equation}

\noindent From \eqref{wave-number} one can define a momentum via $p=\hbar k$ and one can define an energy via $E=\hbar \omega$. Using this $p, E$ and the $k$ defined in \eqref{wave-number} in the relationship $E^2 -p^2 c^2 = m^2 c^4$ leads to the mass given in \eqref{photon-mass} if the frequency $\omega \ge \omega_{plasma}$ \cite{Das08,Claro12}. If $\omega < \omega_{plasma}$ then $m^2 _{\gamma} <0$ and $m_{\gamma}$ becomes imaginary \cite{Weaver74,Afanasev77,Dolgov17,Tarrant21} and the electromagnetic wave will be attenuuated and in the process rapidly heat the plasma. For interplanetary space with a charge density of $\approx 10^7 \frac{1}{m^3}$ \eqref{photon-mass} will give $\omega_{plasma} \simeq 10^5 \frac{1}{Hz}$. For higher plasma densities the plasma frequency will be larger. This is well above the frequency of 100s Hz to 1000s Hz of the gravitational waves we consider. Thus we are in the regime where $\omega < \omega_{plasma}$ then $m^2 _{\gamma} <0$ and $m_{\gamma}$ becomes imaginary. While having a positive, real $m_{\gamma}$ would prohibit the mechanism of gravitational waves producing electromagnetic waves described in section II, having an imaginary mass should still allow this production mechanism of electromagnetic waves. A similar discussion of graviton to photon conversion in the presence of a plasma {\it and} magnetic field can be found in \cite{Dolgov17}. 

\section{Production of sound radiation from gravitational waves} \label{phonon}

There is another possible conversion mechanism in addition to gravitational wave (graviton) to electromagnetic wave (photon) conversion. Since the gravitational wave produced by a CCSN propagates outward in the presence of the dense plasma there exists the possibility of conversion of gravitational waves to sound waves. The presence of the plasma would be associated with an acoustic Lagrangian, $\mathcal{L}_{s}=\mathcal{L}_{s} \left(\rho, \Pi \right)$, and phonon production would occur via the relativistic hydrodynamics of the plasma in the background of the gravitational wave \cite{Visser10,Sabin14,Schutzhold18,Sorge18}. Here the acoustic field is represented by the scalar field, $\Pi$, and the plasma density by $\rho$. The equation of motion for this massless scalar field,  $\Pi(x)$, of the acoustic Lagrangian, $\mathcal{L}_{s}$, is similar to \eqref{eomvarphiA}, and is given by

\begin{equation}
\frac{1}{{\sqrt{-\mathcal{H}}}} \partial_{\mu} \sqrt{-\mathcal{H}} {\mathcal H}^{\mu \nu} \partial_{\nu}\Pi = 0,
\label{eomvarpiA}
\end{equation}

\noindent In \eqref{eomvarpiA} $\mathcal{H}^{\mu \nu}$ is an effective metric which takes into account both the real curved spacetime background, $g^{\mu \nu}$, plus the effects of the plasma/fluid via an acoustic metric, $\mathcal{G}^{\mu \nu}$.  The effective metric for the plane gravitational wave propagating out from the core in the rest frame of the fluid is \cite{Sabin14,Visser10}, 

\begin{equation}
ds^2 = \mathcal{H}_{\mu \nu} dx^{\mu} dx^{\nu} = \frac{1}{c_{s}} \left( \frac{n^{2}_{0} }{\varrho_{0} + p_{0}}\right) \left( -c^{2}_{s}dt^2 + dz^2 + a^2 (t) dx^2 + b^2 (t) dy^2 \right),
\label{GWAmetric}
\end{equation}

\noindent where $c_{s}$ is the speed of sound, $n_{0}$ the number density, $\varrho_{0}$ the energy density, and $p_{0}$ the pressure of the plasma/fluid. Following \cite{Visser10} the pre-factor can be made properly dimensionless, if we ``reinstate appropriate position-independent constants", $ \frac{1}{c_{s}} \left( \frac{n^{2}_{0} }{\varrho_{0} + p_{0}}\right) \rightarrow  \frac{c }{c_{s}} \left( \frac{n^{2}_{0} \varrho_{(p=0)} }{n^{2}_{(p=0)}\left(\varrho_{0} + p_{0}\right)}\right)$. However, ``these extra position-independent constants in the conformal factor carry no useful information and are commonly suppressed.". The constant, overall pre-factor $\frac{1}{c_{s}} \left( \frac{n^{2}_{0} }{\varrho_{0} + p_{0}}\right)$ in $\mathcal{H}_{\mu \nu}$ can be absorbed into the metric by going to the rest frame of the plasma/fluid giving the following equations for the massless scalar field \eqref{eomvarpiA},

\begin{equation}
\frac{1}{{\sqrt{-H}}} \partial_{\mu} \sqrt{-H} H^{\mu \nu} \partial_{\nu}\Pi = 0 ,
\label{eomvarpiA2}
\end{equation}

\noindent but where now the new effective metric, $H^{\mu \nu}$, is

\begin{equation}
ds^2 = H_{\mu \nu} dx^{\mu} dx^{\nu} = -c^2_{s}dt^2 + dz^2 + a^2 (t) dx^2 + b^2 (t) dy^2 .
\label{GWAmetric-2}
\end{equation}

\noindent Note that the components of this effective metric, $a, b$, only depend on $t$ \cite{Sabin14,Schutzhold18}. To determine the solution for the phonon field, $\Pi (z,t)$ we now follow the process used for photons above (more details of this procedure for photons can be found in \cite{Jones16}. We first substitute \eqref{GWAmetric-2} into \eqref{eomvarpiA2} which gives the  phonon field $\Pi (t,z)$ equation as

\begin{widetext}
\begin{equation}
\left( {\frac{1}{a^2} \partial _x^2  + \frac{1}{b^2} \partial _y^2  + \frac{1}{ab}\partial _z \left( {ab} \right)\partial _z  +  \partial _z^2  - c^{-2}_s \partial _t^2  - c^{-2}_s \frac{1}{ab}\partial _t \left( {ab} \right)\partial _t } \right)\Pi  = 0.
\label{AcExpand}
\end{equation}
\end{widetext}

\noindent The oscillatory terms from the metric are of the form $a (t)=1+\varepsilon (t) $ and $b(t)=1-\varepsilon (t)$ with $\varepsilon (t)= h e^{-i\omega t}$, $h$ is the dimensionless strain amplitude and $\omega = k c$ is the frequency of the incoming gravitational wave. We take $\Pi (t,z)$ to be a function of only $t$ and $z$ and not a function of $x$ and $y$. Finally taking into account that, unlike the photon case, here $a,b$ are only a function of $t$ \footnote{As explained in reference \cite{Sabin14} for a fluid one can move to the rest frame of the fluid whereas in vacuum and for photons the best one can do is go to the light front frame.} we simplify equation \eqref{AcExpand} to, 

\begin{widetext}
\begin{equation}
\left( \partial _z^2  - c^{-2}_s \partial _t^2  - c^{-2}_s \frac{1}{ab}\partial _t \left( {ab} \right)\partial _t  \right)\Pi  = 0.
\label{AcExpand1}
\end{equation}
\end{widetext}

\noindent Using $ab=1-\epsilon^2 =1-h^2 e^{-2i\omega t}$ we can expand the last term in \eqref{AcExpand1} and obtain 

\begin{widetext}
\begin{equation}
\left( \partial _z^2  - c^{-2}_s \partial _t^2  +  \frac{2 i h^2 \omega e^{-2i \omega t}}{c_s^2 (1-h^2 e^{-2i \omega t})} \partial_t \right)\Pi  = 0.
\label{AcExpand2}
\end{equation}
\end{widetext}

\noindent We now do separation of variables on \eqref{AcExpand2} by writing the phonon field as $\Pi (z,t) = Z(z) T(t)$. Expanding the denominator $(1-h^2 e^{-2i \omega t})^{-1} \approx 1+ h^2 e^{-2i \omega t} + { \cal O}( h^4)...$ gives for \eqref{AcExpand2}
\begin{widetext}
\begin{equation}
\left( \partial _z^2  - c^{-2}_s \partial _t^2  +  \frac{2 i h^2 \omega e^{-2i \omega t}}{c_s^2} \partial_t \right)\Pi  = 0.
\label{AcExpand2a}
\end{equation}
\end{widetext}

\noindent Inserting $\Pi (z,t) = Z(z) T(t)$ into \eqref{AcExpand2a} and picking a separation constant $K$ gives

\begin{equation}
    \label{Zz}
    \frac{Z''}{Z} = - K^2 \to Z(z) = A e^{\pm i K z}
\end{equation}

\noindent and 

\begin{equation}
    \label{Tt}
    \ddot{T} - 2 i h^2 \omega e^{-2 i \omega t} \dot{T}= - c_s ^2 K^2 T(t) 
\end{equation}

\noindent From the solution in \eqref{Zz} one can see that $K$ is the wave number of the new wave in the fluid. Previewing what is to come we mention that later we will take the amplitude $A$ in \eqref{Zz} to be set by the amplitude of the gravitational wave namely $A=h$. We note that $c_s K = \omega _s$ is the frequency of this wave in the fluid. However one needs $\omega_s = \omega$ i.e. the frequency of the wave in the fluid equals the original gw frequency. This is the usual condition when one goes between different media ({\it e.g.} when light goes from vacuum to glass) where the speed of propagation and the wavelength/wave number change, but the frequency is the same. Therefore $c_s K = \omega_s = \omega = k c$ with $k$ the wave number of the original gravitational wave. One can use this to obtain the relationship between the wave number of the plasma and the wave number of the gravitational wave in vacuum namely $K = c k/c_s$.  This implies $K >k$ and in turn the the wavelength of the sound wave is smaller than the wavelength of the original gravitational wave, $\lambda _{sounds} <\lambda _{gw}$. In fact if the speed of sound in the plasma is small compared to the speed of light, $c \gg c_s$ then  $K \gg k$ and  $\lambda _{sounds} \ll \lambda _{gw}$. Combining all this we find that \eqref{Tt} becomes

\begin{equation}
    \label{Tt1}
    \ddot{T} - 2 i h^2 \omega e^{-2 i \omega t} \dot{T}= - \omega ^2 T(t) 
\end{equation}

\noindent This can be solved to give 

\begin{equation}
    \label{Tt2}
   T(t) = i e^{-x^2/2} x \left[ I_0\left(\frac{x^2}{2} \right) + I_1 \left( \frac{x^2}{2} \right) \right] ~~{\rm where}~~~ x \equiv h e^{-i\omega t} ~,
\end{equation}

\noindent where $I_n (z)$ is the $n^{th}$ modified Bessel function of the first kind. There is another solution to \eqref{Tt1} -- the Meijer G function -- but this blow up for small $x$ i.e. when $h$ is small. Expanding \eqref{Tt2} around $x=0$ (i.e. $h=0$) gives

\begin{equation}
    \label{Tt3}
   T(t) \approx i x - i \frac{x^3}{4} \to i h e^{-i\omega t} = i h e^{- i c k t} = h e^{-ic k t +i \pi/2}~,
\end{equation}

\noindent where we have dropped higher terms in $h$ and written $\omega = k c$ in terms of the initial gravitational wave parameters, and we have moved the pre-factor of $i$ up into the exponential as a phase factor. Putting everything together gives 

\begin{equation}
    \label{pi-sound}
   \Pi (z,t) = Z(z) T(t)= h^2 e^{\pm i K z - i c k t + i \pi /2}~,
\end{equation}

\noindent where we have taken the amplitude of $Z(z)$ as $A=h$ so that in the limit when $c_s \to c$ the present solution for sound waves/phonons matches with the solution in section II for electromagnetic waves/photons. The wave number $K$ of the sound wave is very different from the original gravitational wave wave number, $k$, but this is not unexpected when one goes from one medium (vacuum) to another (plasma).  

Now what we want to obtain is the sound energy flux, $F_{sound}$, from the gw energy flux $F_{gw}=\frac{d E_{gw}}{dt}$. Looking at the discussion above \eqref{emFlux} we used the square of the gravitational Newman-Penrose scalar, $\left| \Psi_4 \right|^2 = 4 k^4 h^2$ , to get $F_{gw} = \frac{c^3h^2 \omega ^2}{16 \pi G}$. Then we used the ratio of electromagnetic flux to gravitational wave flux, namely $\frac{d E_{em}/dt}{d E_{gw}/dt} = \frac{{\left( {\frac{1}{{4\pi }}\left| {\Phi _2 } \right|^2 } \right)}}{{\left( {\frac{1}{{16\pi k^2 }}\left| {\Psi _4 } \right|^2 } \right)}}$, where $\Phi_2$ is the electromagnetic Newman-Penrose scalar. In the case of sound we do not know of a Newman-Penrose scalar for calculating the energy flux due to sound. Therefore in this case we will calculate the sound energy flux using the sound Lagrange density to calculate the energy-momentum tensor for sound and then from this obtain the sound energy flux. The Lagrange density for sound can be written as

\begin{equation}
    \label{l-sound}
    {\cal L}_{s} = \frac{1}{8 \pi} \partial _\mu \Pi \partial ^\mu \Pi ~,
\end{equation}

\noindent which is the scalar field equivalent of the vector, electromagnetic Lagrange density ${\cal L}_{EM} = \frac{1}{16 \pi} F^{\mu \nu} F_{\mu \nu}$. Using the standard procedure the energy-momentum tensor from ${\cal L}_{s}$ is

\begin{equation}
    \label{sound-tensor}
  T^{\mu \nu} _{s} = \frac{1}{4 \pi}\partial ^\mu \Pi \partial ^\nu \Pi - g^{\mu \nu} {\cal L}_{s} = \frac{1}{4 \pi} \partial ^\mu \Pi \partial ^\nu \Pi - \frac{1}{8 \pi}g^{\mu \nu}\partial _\mu \Pi \partial ^\mu \Pi
\end{equation}

The energy flux in the $i^{th}$ direction is now given by $T^{0i} _s$ and since the wave is traveling in the $z$-direction we want $T^{0z}_s = \frac{1}{4 \pi} \partial ^0 \Pi \partial ^z \Pi$. Actually we want the time averaged energy flux which is  

\begin{equation}
    \label{toz}
 \langle T^{0z} _s \rangle = \frac{1}{2} \left(\frac{1}{4 \pi} \partial ^0 \Pi ^* \partial ^z \Pi \right) = \frac{1}{8 \pi c}\partial_t \Pi ^* \partial _z \Pi = \frac{ h^4 k K}{2 \pi}~,
\end{equation}

\noindent where we have used $\Pi (z,t)$ from \eqref{pi-sound} and the fact that $\partial ^0 = \frac{1}{c} \partial _t$. We use $\langle T^{0z} _s \rangle = \frac{ h^4 k K}{2 \pi}$ to replace $\frac{1}{4 \pi} | \Phi_2 |^2$ when we calculate the ratio of sound energy flux to gravitational wave energy flux so that $\frac{d E_{s}/dt}{d E_{gw}/dt} = \frac{\langle T^{0z} \rangle}{\left( {\frac{1}{16\pi k^2 }\left| {\Psi _4 } \right|^2 } \right)} = \frac{2 h^2 K}{k}$. Using all this the flux of sound energy can be written

\begin{eqnarray}
    \label{flux-sound2}
   F_{s} = \frac{d E_{s}}{dt} &=& \frac{\langle T^{0z} \rangle}{\left( {\frac{1}{16\pi k^2 }\left| {\Psi _4 } \right|^2 } \right)} F_{gw} = \left( \frac{2 h^2 K}{k} \right) \left( \frac{c^3 h^2 \omega ^2}{16 \pi G}\right) \nonumber \\
   &=&  \left( \frac{\pi c^3 f^2 h^4}{2 G} \right) \frac{K}{k} = \left( \frac{\pi c^3 f^2 h^4}{2 G} \right) \frac{c}{c_s} ~,
\end{eqnarray}

\noindent where we have used $F_{gw} = \left( \frac{c^3 h^2 \omega ^2}{16 \pi G}\right)$ from section II, $\omega = 2 \pi f$, and $\frac{K}{k} = \frac{c}{c_s}$.

The sound energy flux in \eqref{flux-sound2} was developed here for the first time and is similar to the electromagnetic flux in \eqref{emFlux}, except for an enhancement factor of $\frac{c}{c_s}$. For example, if the speed of sound where $10^4$ smaller \cite{Nambu09} than the speed of light (i.e. $\frac{c}{c_s} = 10^4$) then this enhancement factor would increase the energy pumped into the system by the conversion of gw to sound/phonons by a factor of $10^{4}$, or $10^{44}$ ergs which is still several orders of magnitude too small to contribute to shock recovery. The estimated power production in \eqref{flux-sound2} is first-order and highly dependent on the physical assumptions \cite{Sabin14,Visser10} associated with the acoustic-metric \eqref{GWAmetric} and the speed of sound $\frac{c_s}{c} =\sqrt{\frac{\partial p}{\partial \rho}}$, where $p$ is pressure and $\rho$ is density. These assumptions clearly breakdown in the limit of small ${c_s}$, which has not been previously considered. While we now have a first order solution for production of acoustic radiation by gravitational waves, future work is required to better understand the restrictions imposed by the acoustic-metric formalism e.g., acoustic event horizon \cite{Visser99} and sonic point \cite{Das06,Nambu09}, on the coproduction of acoustic radiation.

The estimated production of sound radiation is highly dependent on the density of the plasma, which could greatly increase the production locally. As in the electromagnetic case of Section \ref{electromagnetic} the production of sound radiation should asymmetrically pump enough energy into the non-burning plasmas to re-ignite fusion. This phonon production in the relativistic hydrodynamics of the plasma is consistent with calculations of dispersion for gravitational waves in the Einstein equations when including the stress energy of the plasma \cite{Madore72,Sacchetti79,Barta18}, which further justifies the inclusion of co-production in models of super-nova remnant evolution.

\section{Comments on the Computational CCSNe models} \label{comments}

In this section we discuss several aspects of the computational models of CCSNe as they relate to the above proposed counterpart electromagnetic/sound radiation production from gravitational waves. 

\subsection{Energetics in one dimensional models}

The timing of when the counterpart electromagnetic or sound radiation is produced is critical in seeing if this gravitational wave co-produced electromagnetic or sound radiation is potentially important in the evolution of the SNR. In this regard one finds the favorable hint that counterpart electromagnetic and sound radiation are very nearly coincident with accepted models of stall and shock recovery \cite{Muller04,Janka07,Janka12}. The coincidence in the timing of the production of counterpart electromagnetic or sound radiation from gravitational waves with the possibility of fusion re-ignition just outside the iron core suggest that these mechanisms for re-ignition of fusion could be important in models of CCSNe and SNR evolution. Because of the wide variation of initial conditions of the progenitor star \cite{Ott16}, and the coincidence with the stall and recovery, it is not possible to determine {\it a priori} the relative importance of the counterpart electromagnetic or sound radiation production and other ``input physics" to progression of the post collapse stellar expansion. Demonstration of the potential importance of the counterpart production mechanisms is only possible if these mechanisms are included in the computational models.

Because of the fourth order dependence of counterpart production on ${\tilde A}$ in \eqref{TotalPowerIntB}, one finds that the mean quadrupole wave amplitude will strongly influence the effect of counterpart radiation production in the supernova recovery. Models of gravitational wave quadrupole amplitudes give amplitudes on the order of centimeters, but can vary from $\sim 5$ cm to $\sim 50$ cm \cite{Muller04,Radice19}. This one order of magnitude change in the quadrupole wave amplitude in \eqref{TotalPowerIntB} can result in large changes in counterpart electromagnetic or sound power production by up to four orders of magnitude. The inclusion of counterpart production in CCSNe simulations would result in models which are very sensitive to the magnitude of the gravitational wave produced in core collapse. Such sensitivity to the gravitational waves coming from core collapse are not found in current models. Current models assume that the gravitational wave propagates freely out from the core and contributes only marginally to the supernova energetics. However, with the strong, fourth order dependence of counterpart electromagnetic and sound radiation production on the gravitational wave quadrupole amplitude, the inclusion of this co-produced radiation in models of stall recovery and stellar expansion would offer a sensitive test for models of gravitational wave production in CCSNe which could be compared with observations.

There are other researchers who have proposed the idea that re-ignition of fusion could play a role in CCSNe. In an early work \cite{Burbidge57} the proposal was made that thermonuclear fusion through gravitational compression could play a role in CCSNe. This proposal was famously abandoned \cite{Janka12A} because the energy produced in the stellar layers is insufficient. The possibility of this conventional thermonuclear mechanism producing a stellar expansion in core-collapse has been recently revived \cite{Kushnir15,Blum16} in 1-dimensional models with highly tuned initial conditions. However, in \cite{Kushnir15} the lower layers of silicon and oxygen are assumed to contribute negligibly to re-ignition of fusion burning. Including the co-produced electromagnetic or sound radiation in these models could significantly effect the burning in the lower layers and hence the evolution of the models such as those proposed in \cite{Kushnir15,Blum16}.

\subsection{Assumptions of three dimensional models}

While inclusion of counterpart electromagnetic and/or sound production in symmetric, 1-dimensional models of core collapse is interesting, the more important applications are likely to be in asymmetric, 3-dimensional models. Re-ignition of fusion in the outer core of the progenitor star, and therefore the equation of state of the stellar material (i.e. temperature, pressure, density) will be asymmetric during the collapse. To investigate the potential influence of the change in local density we briefly review the interaction of the outgoing neutrino radiation with the material of the supernova progenitor. This will demonstrate the potential importance of counterpart electromagnetic or sound production and re-ignition of fusion in the standard neutrino mechanism. The order of magnitude of the neutrino cross section in CCSNe processes is \cite{Bethe85}

\begin{equation}
{\sigma _\nu } \approx 10^{-43} \left[  \frac{E_\nu^2}{1~\rm{MeV^2}} \right] ~\rm{cm^2},
\label{CrossSection}
\end{equation}

\noindent where $E_\nu$ is the energy of the neutrinos in MeV. Using this cross section the neutrino mean free path length is

\begin{equation}
{l_\nu } \approx \frac{m_{u}}{\rho \sigma_\nu },
\label{mfp}
\end{equation}

\noindent where $m_{u} \approx 10^{-24} ~ \rm{g}$ and $\rho$ is the local density. Collecting terms the neutrino mean free path can be written as

\begin{equation}
l_\nu  \approx \left[ \frac{10^{19} ~ \rm{\frac{g}{cm^3}}}{\rho}  \right]  \left[  \frac{1~\rm{MeV^2}}{E_\nu^2} \right] ~\rm{cm} .
\label{mfp2}
\end{equation}

\noindent We see in \eqref{mfp2} that the interaction of the outgoing neutrino radiation with the outer core is determined by the local nucleon density and the neutrino energy. In approximating the neutrino mean free path in \eqref{mfp2} for the outer core, we will assume a CCSN neutrino energy, $E_\nu$, on the order of $10~\rm{MeV}$. The local nucleon density in the outer core is estimated to be on the order of $10^{6}-10^{12}~\rm{\frac{g}{cm^3}}$ \cite{Janka12,Bethe85} resulting in a neutrino mean free path on the order of $10^{5} - 10^{11} ~\rm{cm}$. The equation of state, and therefore the local density, is strongly dependent on nuclear burning. Thus the re-ignition of fusion by the local co-produced radiation would have a strong effect on the local density. This in turn would strongly influence the neutrino mean free path and therefore the effectiveness of neutrinos in exploding the star. Thus, including the effects of the co-produced radiation would make neutrino-driven CCSNe highly dependent on the local power production of the co-produced radiation. The extent of the contribution of this co-produced radiation on neutrino-driven explosions and associated stellar expansion can only be determined through the inclusion of this previously unconsidered physics in core collapse simulations.

\subsection{Electrodynamic Assumptions in Existing Computational Models}

Over the past few decades computer simulations of CCSNe have been greatly improved, but still do not ``explode'' often enough to be consistent with observations of CCSNe \cite{Ott16,Janka17}. The physical processes producing CCSNe are theoretically described by, general relativity, neutrino energetics, nuclear physics, hydrodynamics, and electrodynamics. The simulations of core-collapse are computationally demanding and some physical assumptions and approximations must be made. One method of simplification is to exclude some of the physics of CCSNe in the simulations. For example, many simulations include general relativity, neutrino energetics, nuclear physics, and hydrodynamics, but exclude electrodynamics. Including electrodynamics in the simulation can be computationally demanding. There are three cases generally considered for numerical simulations of CCSNe that include electrodynamics \cite{Bucciantini13,Palenzuela13}.  The first is the isotropic and resistive case, $e^{\mu} =\eta j^{\mu}$, which is essentially the 4-vector form of usual Ohm's Law, ${\bf J} = \sigma {\bf E}$. Here $e^{\mu} = F^{\mu \nu} u_\nu$ is the ``electric" field of an observer co-moving with respect to the fluid. The second case is the inclusion of a mean field dynamo, $e^{\mu} = \xi  b^{\mu} + \eta j^{\mu}$ which is a 4-vector generalization of Ohm's Law to include magnetic fields. Here $b^{\mu} = {\cal F}^{\mu \nu} u_\nu$ is the ``magnetic" field of an observer co-moving with respect to the fluid. The third and final case is the ideal magnetohydrodynamic limit $e^{\mu} = 0$ which in 3-vector form becomes $E^{i} = - \epsilon^{ijk} v_{j} B_{k}$.

If existing simulations of the CCSNe did include more of the physics -- in particular if they included gravitational wave production and full electrodynamics -- the energetics associated with counterpart electromagnetic radiation production would be realized in the computer simulations. However, to date, the most realistic simulations of CCSNe, that include magnetohydrodynamics, impose the ideal magnetohydrodynamic limit, $e^\mu =0$ \cite{Font08}. This assumption eliminates the possibility that these simulations will take into account electromagnetic radiation and in particular the counterpart electromagnetic and phonon radiation proposed here.

\section{Production signatures in CCSN remnants} \label{remnant}

Computational models of core collapse supernovae remnants \cite{Gabler21,Orlando21} are computationally demanding and still in the rather early stage of development. However, these models offer a promising method for observational test of the CCSNe models and the associated physical assumption of these models. These computational models of SNR also have the potential of providing observational evidence of the production of electromagnetic and sound radiation by gravitational waves. While including this co-production in SNR models shares all the challenges of CCSNe models from Section \ref{comments} most computation models of SNR ignore the anisotropies in the gw production. Including the anisotropy in the models for SNR will be essential to any verification of co-production in order to meaningfully compare observations with co-production as one of the physical mechanisms in the models.

Signatures of co-production in SNR and in the computational models would be expected to be distinct from other contributions to the evolution of the expansion due to the unique quadrupole nature \cite{Kotake17} of the production of gravitational waves. This should help distinguish the contribution from re-ignition due to co-production from other process in the explosion and expansion. Even though the energies of the co-production are small compared to other contributions, the unique geometry of the production of gw and subsequent co-production should be evident in computational models provided that the anisotropy of the gw production is included in the model following re-ignition outside the collapsing core.

There are a number of different processes that produce quadrupole moments just after the bounce and generation of gw. These can be divided into processes directly from the core instabilities and the anisotropy in the neutrino emissions. The production of gw by the proto-neutron star can be expressed generically \cite{Kotake17} as,
 
 \begin{equation}
    \label{anisotropic1}
    h_{i,j}\left(\mathbf{X},t \right) = \frac{2G}{c^4 \left| \mathbf{X}\right|} \ddot{I}_{i,j} ~,
\end{equation}

\noindent where $\ddot{I}_{i,j}$ is the the transverse traceless part of the quadrupole moment. Including the production of gw from the core instabilities would provide a signature of counterpart production in SNR models due to the anisotropy in the associated re-ignition of fusion. However, many models of neutrino production of gw assume the far field limit \cite{Muller97,Kotake17} $h_{(+,\times)}\left(\mathbf{X},t \right) \rightarrow h_{(+,\times)}\left( \left| \mathbf{X}\right|, t \right)$ which does not include the anisotropy of the gw production. To model the counterpart production in the SNR this anisotropy must be retained. This is a special problem for gw production by the anisotropy in the neutrino emissions. In a neutrino-driven explosion and subsequent expansion, any signature of the less energetic coproduction of electromagnetic or sound radiation by gw would not be meaningful without taking into account the anisotropies in the models. In this case again including the anisotropy of gw production is critical and the quadrupole nature of gw production  \cite{Vartanyan20} which can be expressed by e.g.,

\begin{equation}
    \label{anisotropic2}
    h_{(+,\times)}\left(\mathbf{X},t \right) = \frac{2G}{c^4 \left| \mathbf{X}\right|}  \int_{0}^{t} dt' \Lambda \left(t' \right) \alpha_{(+,\times)} \left(\alpha, \beta ,t' \right) ~,
\end{equation}

\noindent where $\alpha \in \left[ - \pi, \pi \right] $, $\beta \in \left[ 0, \pi \right] $, $\alpha_{(+,\times)}$ is the polarization dependent ``neutrino emission anisotropy parameter", and $\Lambda \left(t \right)$ is the ``angle-integrated neutrino luminosity". While the energy associated with the neutrino-driven explosion is much greater than the energy associated with co-production, the evolution of the SNR model should show evidence of this coproduction due to the angular dependences in $\alpha_{(+,\times)} \left(\alpha, \beta ,t' \right)$.

Including co-production of sound radiation in computational models of SNR could be comparable to the consideration of $\beta$ decay in existing models \cite{Gabler21}. The additional heating follows nucleosynthesis of heavy elements in the expanding star which would be comparable to processes following reingnition from coproduction. The heating is the order of  $E_{\beta} \sim 10^{48} ~\rm{erg}$ and still far less than the energy required to explode the star. However, the signature of this heating is evident in existing computational models of the SNR, ``These asymmetries particularly manifest themselves in the distribution of the heavy elements, freshly synthesized during the explosion" \cite{Gabler21}. Due to the anisotropy in the production of gw and re-ignition associated with co-production the signatures of this co-production would also be expected to ``manifest themselves".

The sound radiation coproduction was calculated in Section  \ref{phonon} as on the order of  $10^{44} ~\rm{erg}$, but is highly dependent on the local plasma density, which could locally increase the energy production by many orders of magnitude. Including co-production in models of super-nova remnant evolution could provided physically discernible evidence of co-production. The electromagnetic coproduction was calculated in Section \ref{electromagnetic} as on the order of $10^{40} ~\rm{erg}$ which is many orders of magnitude less than the energy estimated for $\beta$ decay. However, this co-production is due to vacuum production, which is independent of the local plasma density. Including co-production in models of super-nova remnant evolution would readily distinguish between contributions from  $\beta$ decay, sound radiation, and vacuum production of electromagnetic radiation. \\

\section{Observational signatures of co-produced radiation}  \label{observation}

While we are currently in an era where multiple 3D simulations per year are possible, the details of the explosion mechanism for core collapse supernova are still debated, with one common agreement that spherical symmetry must be broken at some point to revive the shock which often stalls in simulations. Unable to exactly pin-point the physical requirements of this asymmetry, in the modern era of core-collapse supernova, it has become evident that although the neutrino mechanism appears to be effective at a fundamental level, there are many details that must be better understood. For instance, crucial considerations such as accurately mapping progenitor mass and properties to observable quantities like explosion energy, neutron-star characteristics, nucleosynthesis, morphology, pulsar dynamics, and magnetic field properties, are all relevant in determining the outcome of the collapse (see \cite{Burrows2021} for a review of the current status) because all of these considerations would contribute to the evolution of the explosion and to observable signatures in the remnant.

Our hypothesis regarding the generation of electromagnetic or sound radiation by gravitational waves in core-collapse supernovae could result in observable signatures, particularly at high-energies such as X-rays, which are abundantly produced during the core-collapse process \cite{Ferrand2020}. To obtain precise predictions and better understand the specific observational signatures, it is crucial to conduct detailed computational modeling and analysis within the framework of our proposed conversion mechanisms. Nonetheless, we can outline crude expectations on how this conversion would occur from the production of gravitational radiation. For instance, assuming the conversion of gravitational waves into electromagnetic or sound radiation occurs in bursts, it could manifest as transient X-ray flares. These flares would appear as sudden changes in the X-ray flux which would not be associated with better understood supernova processes. These flux changes would occur on timescales related to the gravitational wave events, and should be evident in the long-term X-ray evolution of the supernova. These transient X-ray flares may be overlooked due to limitations in the sensitivity of current X-ray instruments and, more importantly, the monitoring cadence of X-ray light curves, which often occur over timescales of days to months, or even years \cite{Ross2017}. Furthermore, the co-production of electromagnetic and sound radiation could introduce new components to the X-ray spectrum  that could only be predicted by including co-production in computational models. From an observational perspective, this might manifest as e.g., unusual spectral lines or excess emission at specific energies.

For example, recent James Web Space Telescope (JWST) and {\it Chandra} analysis of the supernova remnant Cassiopeia A found that the unshocked ejecta filaments were connected to Fe-K emission from X-rays \cite{Milisavljevic2024}. This is interesting because the Fe 6.7 keV line is produced when high-energy shocks heat a plasma to temperatures of \( T \sim 10^7 - 10^8 \, \mathrm{K} \), ionizing iron to Fe XXV. It would make sense to see such ionized lines in shocked regions, but it is surprising that they are observed in unshocked regions. It is possible that energy from co-produced radiation could ionize material in these unshocked regions, potentially generating the Fe-K emission in unshocked regions,  although further detailed work is needed to explore this. 

In the next few years, with the advancement in timing and spectral capabilities of future X-ray instruments it should be observationally possible to detect finer features in light curves and spectra. Of particular interest to this study would be the exceptional spectral resolution offered by the X-ray Imaging and Spectroscopy Mission (XRISM, see e.g., \cite{xrism}) and the sensitivity of ATHENA \cite{Decourchelle2013, Jacovich2021}. We believe that now is the opportune time to explore this hypothesis and the potential contributions to the physics underlying CCSNe and CCSNe remnants. \\

\section{Conclusions}

Our rough, first order calculation, of the contribution to the evolution of the supernova remnant from electromagnetic and sonic radiation co-produced by gravitational waves from core-collapse, demonstrates the potential significance of these phenomena in stellar expansion. We also demonstrate that models of core-collapse supernova may be missing a potentially important piece of physics that is generally suppressed in current models of CCSNe and SNR. The point of this work is that computer simulations of core-collapse supernova and models of supernova remnants should consider this process of converting gravitational wave energy into both electromagnetic and sound wave energy, along with more well known mechanisms, and that this would provide evidence of co-production.

Interestingly the idea of the heating of the material in a supernova by phonon production bears a resemblance to mechanisms proposed for heating the solar corona \cite{Aschwanden06}. It has been known for some time that the solar corona, which is several solar radii above the Sun's surface (i.e. the photosphere) is heated to several millions of degrees Kelvin while the photosphere is a much lower temperature of about 6000 K. One of the explanations for this is that sound waves carry energy from the photosphere to the corona and then deposit their energy in the corona as the corona gets too diffuse to propagate the sound waves and their energy further. Here, one of the mechanisms for pumping extra energy into the material of the CCSNe and SNR is for the gravitational waves to produce phonons/sonic radiation which then locally reheats the material to the point where fusion can be re-ignited.

Decay processes and in particular $\beta$ decay of heavy elements synthesized in the explosion of the star were previously demonstrated \cite{Gabler21} to contribute meaningfully to the evolution of supernovae remnants. In large part the potential of observing signatures of $\beta$ decay is due to the anisotropy in the evolution of supernovae remnants. Coproduction of electromagnetic and sound radiation would produce similar decays of heavy elements following local re-ignition associated with coproduction of light and sound by gravitational waves. Including coproduction by gravitational waves in models of the evolution of supernovae remnants will introduce a previously unconsidered physical mechanism in the models, and the signature of the coproduction in these models will provide a method of confirmation of coproduction through associated observations of supernovae remnants. \\

{\noindent {\bf Acknowledgment:}} PJ would like to thank Brennan Hughey for helpful discussions and Quentin Bailey for asking the right questions about existing CCSNe computational models. DS is supported by a 2023-2024 KITP Fellows Award. This research was supported in part by the National Science Foundation under Grant No. NSF PHY-1748958.
\appendix

\end{document}